\documentclass[aps,prd,superscriptaddress,twoside,twocolumn,nofootinbib,showpacs,floatfix]{revtex4-1}
\usepackage{amsmath,amssymb}
\usepackage{graphicx}
\usepackage{subfigure}
\usepackage{color}
\pdfpagewidth=\paperwidth
\pdfpageheight=\paperheight
\allowdisplaybreaks
\newcommand*{\slashed}[1]{{#1\!\!\!/}}
\newcommand*{\hc}{\text{H.\,c.}}

\begin{document}

\title{\boldmath Effects of $N(2000){5/2}^+$ on $\gamma p \to K^+ \Lambda(1405)$}

\author{Yu Zhang}
\affiliation{School of Nuclear Science and Technology, University of Chinese Academy of Sciences, Beijing 101408, China}

\author{Fei Huang}
\email[Corresponding author: ]{huangfei@ucas.ac.cn}
\affiliation{School of Nuclear Science and Technology, University of Chinese Academy of Sciences, Beijing 101408, China}

\date{\today}

\begin{abstract}
The photoproduction reaction of $\gamma p \to K^+\Lambda(1405)$ is investigated based on an effective Lagrangian approach at the tree-level approximation with the purpose of understanding the reaction mechanism and extracting the resonance contents and the associated resonance parameters in this reaction. Apart from the $t$-channel $K$ and $K^\ast$ exchanges, $s$-channel nucleon ($N$) exchange, $u$-channel $\Sigma$, $\Lambda$, and $\Lambda(1405)$ exchanges, and generalized contact term, the exchanges of a minimum number of $N$ resonances in the $s$ channel are taken into account in constructing the reaction amplitudes to describe the experimental data. It is found that by introducing the $N(2000){5/2}^+$ resonance exchange in the $s$ channel, one can reproduce the most recent differential cross-section data from the CLAS Collaboration quite well. Further analysis shows that the cross sections of $\gamma p \to K^+\Lambda(1405)$ at high energies are dominated by the $t$-channel $K$ exchange, while the contributions from the $s$-channel $N$ and $N(2000){5/2}^+$ exchanges are rather significant to the cross sections in the near-threshold energy region. Predictions for the beam and target asymmetries for $\gamma p \to K^+\Lambda(1405)$ are given.
\end{abstract}


\maketitle


\section{Introduction}  \label{sec:Intro}

The study of nucleon resonances ($N^\ast$'s) has always been of great interest in hadron physics, as a deeper understanding of baryon resonances is essential to get insight into the nonperturbative regime of quantum chromodynamics (QCD). Currently, most of our knowledge about the $N^\ast$'s is mainly coming from the $\pi N$ scattering or $\pi$ photoproduction. The productions of mesons other than $\pi$ provide alternative tools to research resonances that couple weakly to $\pi N$ but strongly to other baryon-meson channels. The $\eta N$ and $KY$ ($Y=\Lambda,\Sigma$) channels have been investigated as a first step toward this goal. Recently, the production processes of heavier mesons such as $\eta'$, $\omega$, and $\phi$ have been gaining increasing attention \cite{Huang:2013,Wei2019,Kim2020}. Intense activities have also been performed to study the photoproduction reaction of $K^\ast$ mesons \cite{Ozaki:2010,Kim:2014,Yu:2016,ff3,ac2020c,Wei2020,Zhao:2001jw,Oh:2006in,Kim:2013,ac2018}.

In the present work, we concentrate on the $\gamma p \to K^+\Lambda(1405)$ photoproduction reaction. Since the threshold of $K^+\Lambda(1405)$ is much higher than that of $\pi N$, this reaction is more suitable than $\pi$ production reactions to investigate the $N^\ast$'s in the less-explored higher resonance-mass region. Another advantage of $K^+\Lambda(1405)$ photoproduction in studying $N^\ast$'s is that it acts as an isospin filter, isolating the $N^\ast$'s with isospin $I = 1/2$.

Experimentally, the high-precision differential cross-section data for $\gamma p \to K^+\Lambda(1405)$ in the range of center-of-mass energy $W\approx 2.0$--$2.8$ GeV were published in $2013$ by the CLAS Collaboration at the Thomas Jefferson National Accelerator Facility \cite{data2013}. In Ref.~\cite{data2013}, the total cross-section data obtained by integrating the measured differential cross sections were also reported.

\begin{figure*}[tbp]
\includegraphics[width=0.75\textwidth]{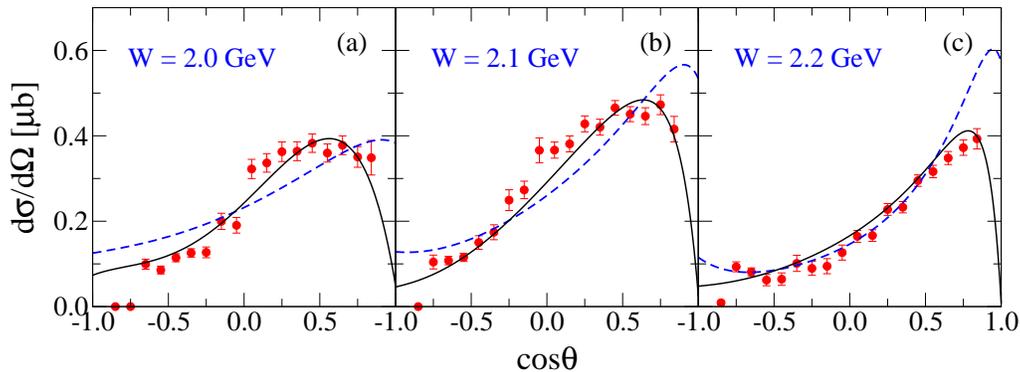}
\caption{(Color online) Status of theoretical description of the differential cross sections for $\gamma p \to K^+ \Lambda(1405)$ at selected center-of-mass energies ($W$) in the near-threshold energy region. The blue dashed lines represent the results from Ref.~\cite{kim}, and the black solid lines denote the results from our present work which is discussed in detail in Sec.~\ref{Sec:results}. The scattered symbols are data from the CLAS Collaboration \cite{data2013}.}
\label{fig:comparison}
\end{figure*}

Theoretically, the CLAS cross-section data for $\gamma p \to K^+\Lambda(1405)$ were analyzed in Refs.~\cite{wangen,kim}. In Ref.~\cite{wangen}, the triangle singularity mechanism for $\gamma p \to K^+\Lambda(1405)$ was investigated. There, assuming that the transition of $\gamma p\to K^\ast \Sigma$ is dominated by an $N(2030){1/2}^-$ resonance with a width of $300$ MeV, it is reported that the mechanism where the intermediate $K^\ast$ decays to $K^+\pi$ and the $\pi\Sigma$ merge to form the $\Lambda(1405)$ can produce a peak around $W=2110$ MeV and has its largest contribution around $\cos\theta=0$, leading to a good reproduction of the cross-section data in the energy range $W\approx 2.0$--$2.2$ GeV. Nevertheless, a detailed analysis of the cross-section data for $\gamma p\to K^{\ast +}\Sigma^0$ and $\gamma p\to K^{\ast 0}\Sigma^+$ performed in Ref.~\cite{ac2018} shows that the required resonance contribution in $K^\ast \Sigma$ photoproduction is from the $\Delta(1905){5/2}^+$ resonance exchange and the cross sections of $\gamma p\to K^\ast \Sigma$ are rather small around $W\approx 2.1$ GeV. In this case, the effects of the $K^\ast \Sigma$ intermediate state on the $\gamma p \to K^+\Lambda(1405)$ reaction, which is regarded as a triangle singularity mechanism in Ref.~\cite{wangen}, might need to be further inspected. In Ref.~\cite{kim}, the CLAS cross-section data for $\gamma p \to K^+\Lambda(1405)$ in the energy range $W\approx 2.0$--$2.8$ GeV were analyzed within an effective Lagrangian approach. There, five nucleon resonances are considered, namely, the $N(2000){5/2}^+$, $N(2100){1/2}^+$, $N(2030){1/2}^-$, $N(2055){3/2}^-$, and $N(2095){3/2}^-$ resonances, with the former two taken from the review of the Particle Data Group (PDG) \cite{PDG} and the latter three taken from the quark model calculation of Refs.~\cite{capstick92,capstick98}. A common decay width $\Gamma = 300$ MeV is set for all these five nucleon resonances, and the resonance hadronic and electromagnetic couplings are determined by the corresponding resonance decay amplitudes given by either PDG \cite{PDG} or the quark model calculations \cite{capstick92,capstick98}. The results of Ref.~\cite{kim} show that the contributions from the $N(2000){5/2}^+$ and $N(2100){1/2}^+$ resonances are important to the total cross sections, while for differential cross sections, considerable contributions from the $N(2000){5/2}^+$ and $N(2100){1/2}^+$ resonances are only seen at the energy point of $W\approx 2.1$ GeV. We mention that in Ref.~\cite{Nam2017} the photo- and electroproduction of $\Lambda(1405)$ via $\gamma^{(\ast)}p\to K^+ \pi^+ \Sigma^-$ was investigated.

The work of Ref.~\cite{kim} presents so far the most detailed analysis of the available cross-section data for $\gamma p \to K^+\Lambda(1405)$ in the energy region of $W\approx 2.0$--$2.8$ GeV covered by the CLAS experiments \cite{data2013}. It describes the data on total cross sections qualitatively well in the whole energy region considered, and the differential cross-section data have also been qualitatively described in the energy region of $W>2.3$ GeV. However, for $W<2.3$ GeV where the nucleon resonances are relevant, obvious discrepancies are seen. As an illustration, in Fig.~\ref{fig:comparison} we show a comparison of the differential cross sections calculated in Ref.~\cite{kim} with the CLAS data \cite{data2013} at three selected energy points near the $K^+\Lambda(1405)$ threshold. There, the blue dotted lines represent the theoretical results from Ref.~\cite{kim}, and the black solid lines correspond to the results from our present work which is discussed in detail in Sec.~\ref{Sec:results}. The scattered symbols are data from the CLAS Collaboration \cite{data2013}. It is obviously seen from Fig.~\ref{fig:comparison} that there is still much room for improvement in the differential cross-section results of Ref.~\cite{kim}. One naturally expects that a better reproduction of the data, especially in the near-threshold energy region where the nucleon resonance is relevant, is necessary and essential to understand better the reaction mechanism and determine better the resonance contents and the associated resonance parameters in the $\gamma p \to K^+\Lambda(1405)$ reaction.

In the present work, we analyze the available differential cross-section data from the CLAS Collaboration \cite{data2013} for the $\gamma p \to K^+\Lambda(1405)$ reaction within an effective Lagrangian approach at the tree-level  approximation. The paper is organized as follows. In Sec.~\ref{Sec:forma}, we briefly introduce the framework of our theoretical model. In Sec.~\ref{Sec:results}, we show the results of our theoretical calculations, and discuss the reaction dynamics, the resonance contents, and their associated parameters in the $\gamma p \to K^+\Lambda(1405)$ reaction. Finally, we give a brief summary and conclusions in Sec.~\ref{sec:summary}.

\section{Formalism}  \label{Sec:forma}

\begin{figure}[tbp]
\subfigure[~$s$ channel]{
\includegraphics[width=0.45\columnwidth]{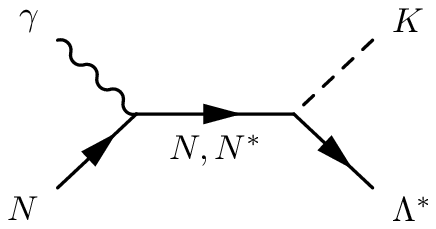}}  {\hglue 0.4cm}
\subfigure[~$t$ channel]{
\includegraphics[width=0.45\columnwidth]{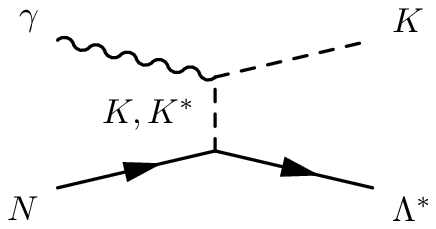}} \\[6pt]
\subfigure[~$u$ channel]{
\includegraphics[width=0.45\columnwidth]{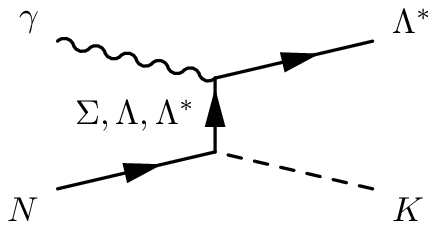}} {\hglue 0.4cm}
\subfigure[~Interaction current]{
\includegraphics[width=0.45\columnwidth]{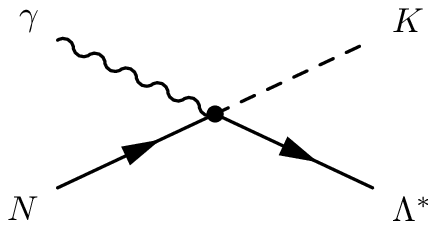}}
\caption{Generic structure of the amplitudes for $\gamma p\to K^+ \Lambda(1405)$. Time proceeds from left to right. The symbols $\Lambda^\ast$ denote $\Lambda(1405)$.}
\label{FIG:feymans}
\end{figure}

Following a full field theoretical approach of Refs.~\cite{Haberzettl:1997,Haberzettl:2006}, the full photoproduction amplitudes for $\gamma p \to K^+ \Lambda(1405)$ can be expressed as
\begin{equation}
M^{\mu} = M^{\mu}_s + M^{\mu}_t + M^{\mu}_u + M^{\mu}_{\rm int},  \label{eq:amplitude}
\end{equation}
with $\mu$ being the Lorentz index of the incoming photon. The first three terms, $M^{\mu}_s$, $M^{\mu}_t$, and $M^{\mu}_u$, stand for the $s$-, $t$-, and $u$-channel pole diagrams, respectively, with $s$, $t$, and $u$ being the corresponding Mandelstam variables of the internally exchanged particles. They arise from the photon attaching to the external particles of the $KN\Lambda(1405)$ interaction vertex. The last term, $M^{\mu}_{\rm int}$, stands for the interaction current which arises from the photon attaching to the internal structure of the $KN\Lambda(1405)$ interaction vertex. All those four terms in Eq.~(\ref{eq:amplitude}) are diagrammatically depicted in Fig.~\ref{FIG:feymans}.

As represented in Fig.~\ref{FIG:feymans}, the following contributions are considered in our calculation: $N$ and $N^\ast$ exchanges in the $s$ channel, $K$ and $K^\ast$ exchanges in the $t$ channel, $\Sigma$, $\Lambda$,  and $\Lambda(1405)$ exchanges in the $u$ channel, and the interaction current. The amplitudes for the $s$-, $t$-, and $u$-channel hadron exchanges can be constructed straightforwardly by evaluating the corresponding Feynman diagrams. The amplitudes for the interaction current cannot be calculated exactly since they obey highly nonlinear equations. Following Refs.~\cite{Haberzettl:1997,Haberzettl:2006,Huang:2012,Huang:2013}, we model the interaction current as
\begin{equation}
\mathcal{M}_{\rm int}^\mu = \Gamma_{\Lambda^\ast N K} C^\mu,
\end{equation}
where $\Gamma_{\Lambda^\ast N K}=g_{\Lambda^\ast NK}$ stands for the $\Lambda(1405) N K$ vertex function given by the Lagrangian of Eq.~(\ref{eq:L_LstNK}), and $C^\mu$ is an auxiliary current introduced to ensure that the full photoproduction amplitudes of Eq.~(\ref{eq:amplitude}) satisfy the generalized Ward-Takahashi identity (WTI) and thus are fully gauge invariant. Following Refs.~\cite{ff3,ac2018,Wei2019,ac2020}, we choose $C^\mu$ for $\gamma p \to K^+\Lambda(1405)$ as
\begin{equation}
C^\mu =  - e \frac{f_t-\hat{F}}{t-q^2}  (2q-k)^\mu - e \frac{f_s-\hat{F}}{s-p^2} (2p+k)^\mu,
\end{equation}
with
\begin{equation} \label{eq:Fhat}
\hat{F} = 1 - \hat{h} (1 - f_t) (1 -  f_s) .
\end{equation}
Here $q$, $k$, and $p$ are four-momenta of the outgoing $K$, incoming photon, and incoming $N$, respectively; $f_t$ and $f_s$ are form factors attached to the hadronic vertices for $t$-channel $K$ exchange and $s$-channel $N$ exchange, respectively; and $\hat{h}$ is an arbitrary function going to unity in the high-energy limit and set to be $\hat{h}=1$ for simplicity in the present work.

In the rest of this section, we present the effective Lagrangians, the resonance propagator, and the phenomenological form factors employed in the present work.

\subsection{Effective Lagrangians} \label{Sec:Lagrangians}

For the sake of simplicity, we use the character $\Lambda^\ast$ to denote $\Lambda(1405)$ in this section.

The electromagnetic interaction Lagrangians required to calculate the nonresonant Feynman diagrams read
\begin{eqnarray}
\mathcal{L}_{\gamma NN}   &=& -e \bar N \left[ \left(\hat{e} \gamma^\mu - \frac{\hat{\kappa}_N}{2M_N}\sigma^{\mu\nu}\partial_\nu \right) A_\mu \right] N,      \label{rNN}     \\[6pt]
\mathcal{L}_{\gamma KK}   &=& ie\left[K^+\left(\partial_\mu K^-\right) - K^-\left(\partial_\mu K^+\right)\right]A^\mu,   \label{L_rKK}    \\[6pt]
\mathcal{L}_{\gamma KK^\ast} &=& e \frac{g_{\gamma KK^\ast}}{M_K}\epsilon^{\mu\nu\alpha\beta} \left(\partial_\mu A_\nu\right) \left(\partial_\alpha K  \right) K^{\ast}_\beta,      \label{L_rKKS}     \\[6pt]
\mathcal{L}_{\gamma\Lambda^\ast\Lambda^\ast}   &=& e\frac{\kappa_{\Lambda^\ast}}{2M_N} \bar{\Lambda}^\ast \sigma_{\mu\nu} \left(\partial^\nu A^\mu\right) \Lambda^\ast,      \label{L_rLsLs}      \\[6pt]
\mathcal{L}_{\gamma\Lambda\Lambda^\ast}   & =& e\frac{\kappa_{\Lambda^\ast \Lambda}}{2M_N}\bar{\Lambda}\gamma_5\sigma_{\mu\nu} \left(\partial^\nu A^\mu\right) \Lambda^\ast + \hc,  \label{L_rLLs} \\[6pt]
\mathcal{L}_{\gamma\Sigma\Lambda^\ast}   &=& e \frac{\kappa_{\Lambda^\ast \Sigma}}{2M_N}\bar{\Sigma}^0\gamma_5\sigma_{\mu\nu} \left(\partial^\nu A^\mu\right) \Lambda^\ast + \hc,   \label{L_rSLs}
\end{eqnarray}
where $e$ is the elementary charge unit and $\hat{e}$ stands for the charge operator; $\hat{\kappa}_N$ is the anomalous magnetic moment for nucleon and its value is $1.793$ for proton; the electromagnetic coupling $g_{\gamma KK^\ast} = -0.413$ is taken from Ref.~\cite{ff3}, determined by the radiative decay width of $K^\ast\to K \gamma$ with the sign inferred from the SU(3) flavor symmetry in conjunction with the vector meson dominance assumption; $\kappa_{\Lambda^\ast} = 0.44$, $\kappa_{\Lambda^\ast \Lambda} = -0.43$, and $\kappa_{\Lambda^\ast \Sigma} = 0.61$ are anomalous magnetic moments for the $\Lambda(1405)$, $\Lambda(1405)\to \Lambda\gamma$ transition, and $\Lambda(1405)\to \Sigma^0\gamma$ transition, respectively, taken from Ref.~\cite{kim}. Originally, $\kappa_{\Lambda^\ast \Lambda}$ and $\kappa_{\Lambda^\ast \Sigma}$ are calculated from an isobar model of Ref.~\cite{lam} to match the $K^- p$ atom data of Ref.~\cite{sig}. For $\kappa_{\Lambda^\ast}$, as there are no data, its value is taken from the quark model as done in Ref.~\cite{lams}. 

The effective Lagrangians for meson-baryon interactions are
\begin{eqnarray}
\mathcal{L}_{\Lambda NK} &=& - \frac{g_{\Lambda NK}}{2M_N} \bar{\Lambda} \gamma_5 \gamma^\mu \left( \partial_\mu K \right) N + \hc, \\[6pt]
\mathcal{L}_{\Sigma NK} &=& - \frac{g_{\Sigma NK}}{2M_N} \bar{\Sigma} \gamma_5 \gamma^\mu \left( \partial_\mu K \right) N + \hc,   \\[6pt]
\mathcal{L}_{\Lambda^\ast NK} &=& i g_{\Lambda^\ast NK} \bar{\Lambda}^\ast KN + \hc, \label{eq:L_LstNK} \\[6pt]
\mathcal{L}_{\Lambda^\ast NK^\ast} &=& -g_{\Lambda^\ast NK^\ast} \bar{\Lambda}^\ast \gamma_5\gamma^\mu K^{\ast}_\mu N + \hc
\end{eqnarray}
Note that following Ref.~\cite{ff3}, a pure pseudovector coupling is chosen for the $\Lambda NK$ vertex. By use of the SU(3) flavor symmetry, the same choice is made for the $\Sigma NK$ vertex. The coupling constants $g_{\Lambda NK} = -13.99$ and $g_{\Sigma NK} = 2.692$ are taken from Refs.~\cite{ff3} and \cite{ac2018}, respectively, determined by the SU(3) flavor symmetry:
\begin{eqnarray}
g_{\Lambda NK} &=& - \frac{3\sqrt{3}} {5} g_{NN\pi},  \\[6pt]
g_{\Sigma NK} &=& \frac{1}{5} g_{NN\pi},
\end{eqnarray}
with the empirical value $g_{NN\pi} = 13.46$. The coupling constants $g_{\Lambda^\ast NK}$ and $g_{\Lambda^\ast N K^\ast}$ are treated as  fit parameters in the present work due to the scarcity of experimental information.

The Lagrangians for resonance $N(2000){5/2}^+$ electromagnetic and hadronic couplings read \cite{Rushbrooke1965}
\begin{eqnarray}
{\cal L}_{RN\gamma}^{5/2+}  &= & e\frac{g_{RN\gamma}^{(1)}}{\left(2M_N\right)^2}\bar{R}_{\mu \alpha}\gamma_\nu \left(\partial^{\alpha} F^{\mu \nu}\right)N \nonumber \\
&& +\, ie\frac{g_{RN\gamma}^{(2)}}{\left(2M_N\right)^3}\bar{R}_{\mu \alpha} \left(\partial^\alpha F^{\mu \nu}\right)\partial_\nu N + \hc,  \label{eq:L_RNr}  \\[6pt]
{\cal L}_{R\Lambda^\ast K}^{5/2+} &=& i\frac{g_{R\Lambda^\ast K}}{M^2_K} \bar{\Lambda}^\ast \left( \partial^\mu\partial^\nu K \right)  R_{\mu\nu}  + \hc,   \label{eq:L_RLK}
\end{eqnarray}
where $R$ designates the $N$ resonance, and the superscripts of ${\cal L}_{RN\gamma}$ and ${\cal L}_{R\Lambda^\ast K}$ denote the spin and parity of the resonance $R$. The field-tensor $F^{\mu \nu}$ reads
\begin{equation}
F^{\mu \nu} \equiv \partial^\mu A^\nu - \partial^\nu A^\mu.
\end{equation}
The products of the coupling constants $g_{RN\gamma}^{(i)}g_{R\Lambda^\ast K}$ $(i=1,2)$, which are relevant to the amplitudes of $\gamma p\to K^+\Lambda(1405)$, are treated as fit parameters in the present work due to the lack of experimental information on the resonance $N(2000){5/2}^+$ decay.

\subsection{Resonance Propagator} \label{prop}

Following Refs.~\cite{Behrends:1957,Fronsdal:1958,Zhu:1999}, the prescription of the propagator for resonance with spin-$5/2$ reads
\begin{eqnarray}
S_{5/2}(p) &=&  \frac{i}{\slashed{p} - M_R + i \Gamma_R/2} \,\bigg[ \, \frac{1}{2} \big(\tilde{g}_{\mu \alpha} \tilde{g}_{\nu \beta} + \tilde{g}_{\mu \beta} \tilde{g}_{\nu \alpha} \big)  \nonumber \\
&& -\, \frac{1}{5}\tilde{g}_{\mu \nu}\tilde{g}_{\alpha \beta}  + \frac{1}{10} \big(\tilde{g}_{\mu \alpha}\tilde{\gamma}_{\nu} \tilde{\gamma}_{\beta} + \tilde{g}_{\mu \beta}\tilde{\gamma}_{\nu} \tilde{\gamma}_{\alpha}  \nonumber \\
&& +\, \tilde{g}_{\nu \alpha}\tilde{\gamma}_{\mu} \tilde{\gamma}_{\beta} +\tilde{g}_{\nu \beta}\tilde{\gamma}_{\mu} \tilde{\gamma}_{\alpha} \big) \bigg],   \label{eq:prop-5/2}
\end{eqnarray}
with
\begin{eqnarray}
\tilde{g}_{\mu \nu} &=& - g_{\mu \nu} + \frac{p_{\mu} p_{\nu}}{M_R^2},  \\[6pt]
\tilde{\gamma}_{\mu} &=& \gamma^{\nu} \tilde{g}_{\nu \mu} = -\gamma_{\mu} + \frac{p_{\mu}\slashed{p}}{M_R^2},
\end{eqnarray}
where $M_R$ and $\Gamma_R$ are, respectively, the mass and width of resonance $R$, and $p$ is the resonance four-momentum.

We mention that the prescriptions for vertices and propagators of high spin resonances are not unique. In Refs.~\cite{Vrancx2011,David1996,Pascalutsa1999} prescriptions different from these in Eqs.~(\ref{eq:L_RNr}), (\ref{eq:L_RLK}), and (\ref{eq:prop-5/2}) are discussed.

\subsection{Form Factors} \label{ff}

In the present work, each hadronic vertex obtained from the Lagrangians given in Sec.~\ref{Sec:Lagrangians} is accompanied with a phenomenological form factor to parametrize the structure of the hadrons and to normalize the behavior of the production amplitude. Following Refs.~\cite{ff3,ac2018}, for intermediate baryon exchange we use the following form factor:
\begin{equation}
f_{s(u)}\left(p^2\right) = \left( \frac{\Lambda_B^4}{\Lambda_B^4+\left(p^2-M_B^2\right)^2} \right)^2 ,
\end{equation}
with $p$, $M_B$, and $\Lambda_B$ denoting the four-momentum, the mass, and the cutoff mass for the exchanged baryon $B$, respectively. To reduce the number of fit parameters, we use the same cutoff mass for the $u$-channel $\Sigma$, $\Lambda$, and $\Lambda(1405)$ exchanges. As both the $s$-channel $N$ and resonance exchanges make considerable contributions in this work, their cutoff masses are fitted respectively. For intermediate meson exchange, we use the following form factor:
\begin{equation}
f_t\left(q^2\right) =  \left(\frac{\Lambda_M^2-M_M^2}{\Lambda_M^2-q^2}\right)^2,
\end{equation}
with $q$, $M_M$, and $\Lambda_M$ representing the four-momentum, the mass, and the cutoff mass for the exchanged meson $M$, respectively. To reduce the number of fit parameters, we use the same cutoff mass for the $t$-channel $K$ and $K^\ast$ exchanges.

Note that the gauge-invariance feature of our photoproduction amplitude is independent of any specific form of the form factors attached to the hadronic vertices of the $s$-, $t$-, and $u$-channel interacting diagrams.

\section{Results and discussion}   \label{Sec:results}

As mentioned in Sec.~\ref{sec:Intro}, the work of Ref.~\cite{kim} presents so far the most detailed analysis of the CLAS cross-section data for the $\gamma p \to K^+\Lambda(1405)$ photoproduction reaction. Although it describes the differential cross-section data above $W>2.3$ GeV qualitatively well, there is still much room for improvement for $W<2.3$ GeV where the nucleon resonances are relevant, as illustrated in Fig.~\ref{fig:comparison}.

\begin{table}[tbp]
\caption{\label{Table:para} Fitted values of adjustable model parameters. See Sec.~\ref{Sec:forma} for their definitions. The asterisks to the right of resonance name denote the overall status of this resonance rated by PDG \cite{PDG}. }
\begin{tabular*}{\columnwidth}{@{\extracolsep\fill}lr}
\hline\hline
$g_{\Lambda^\ast NK}$  			&   $2.43\pm0.06$ 	\\
$g_{\Lambda^\ast NK^\ast}$  		&   $2.04\pm 0.57$ 	\\
$\Lambda_N$ (MeV)   			&   $1641\pm 18$   	 \\
$\Lambda_{\Sigma, \Lambda, \Lambda^\ast}$  (MeV)  	&   $1898 \pm 42$  \\
$\Lambda_{K^\ast,K}$ (MeV)  		&   $1261 \pm 19$ 	\\[2pt]
\hline
$N(2000)5/2^+$ 			      &  $\ast\ast$ 	\\
$M_R$  (MeV)  					&    $1913\pm 7 $   	\\
$\Gamma_R$  (MeV)  			&    $471\pm 81 $ 	\\
$\Lambda_{R}$ (MeV)  			&    $803 \pm 17$ 	\\
$g^{(1)}_{RN\gamma}g_{R\Lambda^\ast K}$ 	&    $ 40.10 \pm 9.88  $    \\
$g^{(2)}_{RN\gamma}g_{R\Lambda^\ast K}$ 	&    $ -31.27 \pm 7.93  $   \\
\hline\hline
\end{tabular*}
\end{table}

\begin{figure*}[tbp]
\includegraphics[width=0.75\textwidth]{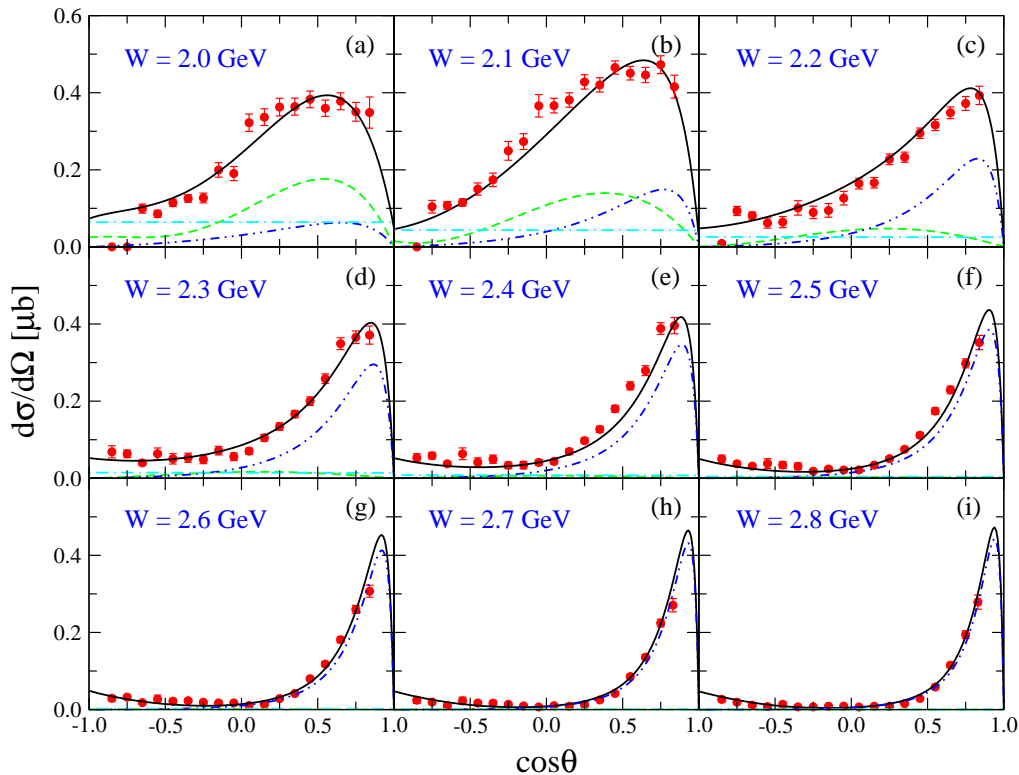}
\caption{(Color online) Differential cross sections for $\gamma p \to K^+ \Lambda(1405)$ as a function of $\cos\theta$.  The black solid lines represent the results from the full calculation. The green dashed, blue dash-double-dotted, and cyan dash-dotted lines represent the individual contributions from the $s$-channel $N(2000){5/2}^+$ exchange, the $t$-channel $K$ exchange, and the $s$-channel $N$ exchange, respectively. The scattered symbols denote the data from the CLAS Collaboration \cite{data2013}. }
\label{fig:5}
\end{figure*}

In the present work, we perform a detailed analysis of the available differential cross-section data from the CLAS Collaboration \cite{data2013} for $\gamma p \to K^+\Lambda(1405)$ within an effective Lagrangian approach at the tree-level approximation. The Feynman diagrams we considered are shown in Fig.~\ref{FIG:feymans}. For the $s$-channel interaction, in addition to the $N$ exchange, we introduce as few as possible nucleon resonances to describe the data. 

If we do not introduce any nucleon resonance in the $s$ channel, it is found that the differential cross-section data cannot be satisfactorily described, especially in the near-threshold energy region. We then try to introduce one nucleon resonance in constructing the production amplitudes for this reaction. We test one by one all of the nucleon resonances near the $K^+\Lambda(1405)$ threshold. If the tested resonance is rated as a four-star or three-star resonance in the PDG \cite{PDG}, we fix the resonance mass and width by the PDG values, and determine the resonance hadronic and electromagnetic coupling constants by the corresponding resonance decay amplitudes or branching ratios advocated in PDG \cite{PDG}. However, if the tested resonance is rated as a two-star or one-star resonance in the PDG \cite{PDG}, which means that this resonance is not well established and their experimental information is scarce, we treat the resonance mass, width, and coupling constants as fit parameters to be determined by fitting the available differential cross-section data. After a lot of tests, it is found that the CLAS data on angular distributions for $\gamma p \to K^+\Lambda(1405)$ can be well described by including a ${5/2}^+$ resonance. In the most recent version of PDG \cite{PDG}, there are two ${5/2}^+$ resonances, i.e., $N(1860)$ and $N(2000)$, both rated as two-star resonances. Although our fitted mass and width of the ${5/2}^+$ resonance are comparable to the corresponding values of both $N(1860)$ and $N(2000)$ as indicated in PDG \cite{PDG}, we prefer to identify the needed ${5/2}^+$ resonance as $N(2000)$ under the consideration that the $N(2000)$ has a branching ratio to $\Lambda K^\ast$ while there is no information about the decay of $N(1860)$ to any strange baryon-meson channels in PDG \cite{PDG}.\footnote{Note that the $N(2000){5/2}^+$ resonance has also been found to be rather important for reproducing the data on spin density matrix elements for $\gamma p \to K^{\ast +}\Lambda$ \cite{Wei2020}.} By considering any one of the other nucleon resonances instead of the ${5/2}^+$ one, the resulting $\chi^2$ will be much larger, indicating a rather bad fitting quality, and thus such a fit is not considered an acceptable one. If two or more nucleon resonances are taken into account, the fitting quality will be improved a little bit, but in this case one does not get any conclusive arguments in regards to the resonance contents and parameters, as there are too many solutions with similar $\chi^2$ which cannot be distinguished by the available cross-section data alone. We postpone such analysis with the introduction of two or more nucleon resonances to future work when more data on other observables for this reaction become available.

As mentioned above, the CLAS differential cross-section data for $\gamma p \to K^+\Lambda(1405)$ can be well reproduced in our model with the inclusion of the $N(2000){5/2}^+$ resonance. The $N(2000){5/2}^+$ is rated as a two-star resonance in the PDG \cite{PDG}, which means that it is not well established yet and we do not have enough information for its parameters. Therefore, in the present work, we treat its mass $M_R$, width $\Gamma_R$, and the products of its hadronic and electromagnetic coupling constants $g_{RN\gamma}^{(i)}g_{R\Lambda^\ast K}$ $(i=1,2)$ as fit parameters. The fitted values of all the adjustable model parameters are listed in Table~\ref{Table:para}. In this table, the uncertainties of the values of fit parameters are estimated from the uncertainties (error bars) associated with the fitted experimental differential cross-section data. The values of $\Lambda_{K^\ast,K}$, $\Lambda_N$, and $\Lambda_{\Sigma, \Lambda, \Lambda^\ast}$ indicate the cutoff parameters for the $t$-channel $K$ and $K^\ast$ exchanges, the $s$-channel $N$ exchange, and the $u$-channel $\Sigma$, $\Lambda$, and $\Lambda(1405)$ exchanges, respectively. The asterisks to the right of resonance name denote the overall status of this resonance rated by the PDG \cite{PDG}.

The theoretical results of the differential cross sections for $\gamma p \to K^+\Lambda(1405)$ corresponding to the parameters listed in Table~\ref{Table:para} are shown in Fig.~\ref{fig:5}. There, the green dashed, blue dash-double-dotted, and cyan dash-dotted lines represent the individual contributions from the $s$-channel $N(2000){5/2}^+$ resonance exchange, $t$-channel $K$ exchange, and $s$-channel $N$ exchange, respectively. The individual contributions from other terms are too small to be clearly seen with the scale used, and thus they are not plotted in Fig.~\ref{fig:5}. The scattered symbols represent the data from the CLAS Collaboration \cite{data2013}. In total ${\rm ND} = 158$ data points are fitted and the resulted $\chi^2/{\rm ND}=3.44$. One sees from Fig.~\ref{fig:5} that the overall agreement of our theoretical differential cross-section results with the corresponding data is very good. In particular, our description of the data is much better than that in Ref.~\cite{kim} in the energy region of $W<2.3$ GeV (see Fig.~\ref{fig:comparison} for a direct comparison). In the high-energy region, it is seen that the differential cross sections are dominated by the contributions from the $t$-channel $K$ exchange, which results in a sharp rise at forward angles. In the low-energy region, the contributions from the $t$-channel $K$ exchange are significant at forward angles, but they are less important than those in the high-energy region. Considerable contributions from the $s$-channel $N$ and $N(2000){5/2}^+$ exchanges are seen at the intermediate and backward angles in the near-threshold energy region. Although our theoretical differential cross-section results are  in an overall good agreement with the data over the entire energy region considered, one still sees small deviations of our predicated angular distributions from the data around $\cos\theta\approx 0$ at $W=2100$ MeV. Further investigations will be done in our future work to check whether these deviations are coming from the coupled-channel effects of the intermediate $K^\ast \Sigma$ states which are claimed as the role of triangle singularity in Ref.~\cite{wangen}.

\begin{figure}[tbp]
\includegraphics[width=0.9\columnwidth]{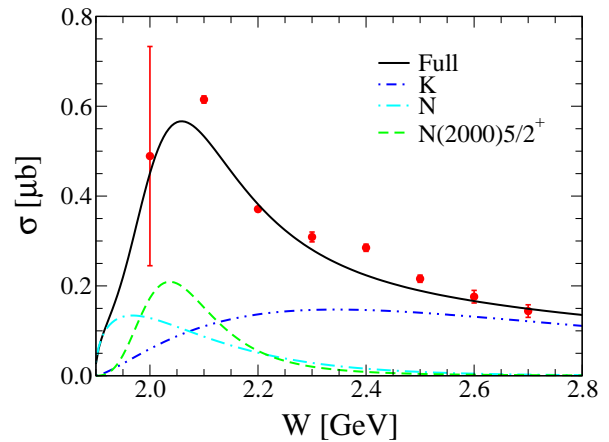}
\caption{(Color online) Total cross sections with dominant individual contributions for $\gamma p \to K^+\Lambda(1405)$ as a function of the center-of-mass energy $W$. The notations are the same as in Fig.~\ref{fig:5}. The data are from the CLAS Collaboration \cite{data2013} but not included in the fit. }
\label{fig:sig}
\end{figure}

Figure \ref{fig:sig} shows our predicted total cross sections (black solid lines) together with individual contributions from the $t$-channel $K$ exchange (blue dash-double-dotted line), $s$-channel $N$ exchange (cyan dash-doted line), and $s$-channel $N(2000){5/2}^+$ resonance exchange (green dashed line). With the present scale used, the contributions from other terms are too small to be plotted in this figure. Note that the total cross-section data \cite{data2013} are not included in our fit. Nevertheless, one sees from Fig.~\ref{fig:sig} that our theoretical results of the total cross sections are in qualitative agreement with the CLAS data over the entire energy region considered. The contributions from the $t$-channel $K$ exchange are seen to dominate the total cross sections at high energies. The $s$-channel $N$ and $N(2000){5/2}^+$ resonance exchanges are seen to have considerable contributions to the total cross sections at low energies, and the $N(2000){5/2}^+$ resonance exchange is responsible for the bump structure exhibited by the CLAS total cross-section data around $W\approx 2.1$ GeV. One may observe that our theoretical total cross sections underestimate the data at the energy points of $W=2.1$, $2.3$, $2.4$, and $2.5$ GeV, although our theoretical differential cross sections at these energies agree quite well with the corresponding data. In this regard, it should be mentioned that the CLAS total cross-section data---which are obtained by integrating the measured differential cross sections---may suffer from the limited angular acceptance of the CLAS detector \cite{data2013}. Thus, one does not need to worry too much about the modest discrepancies between the theoretical total cross sections and the data.

\begin{figure}[tbp]
\includegraphics[width=\columnwidth]{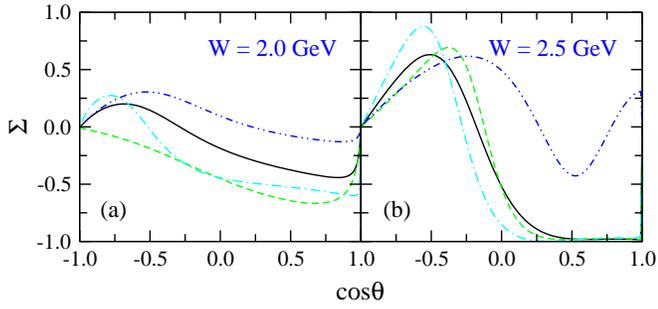}
\caption{(Color online) Photon beam asymmetries as functions of $\cos\theta$ for $\gamma p \to K^+\Lambda(1405)$. The black solid lines represent the results from the full calculation. The green dashed, blue dash-double-dotted, and cyan dash-dotted lines represent the results obtained by switching off the individual contributions of the $s$-channel $N(2000){5/2}^+$ exchange, the $t$-channel $K$ exchange, and the $s$-channel $N$ exchange, respectively, from the full model. }
\label{fig:beam}
\end{figure}

\begin{figure}[tbp]
\includegraphics[width=\columnwidth]{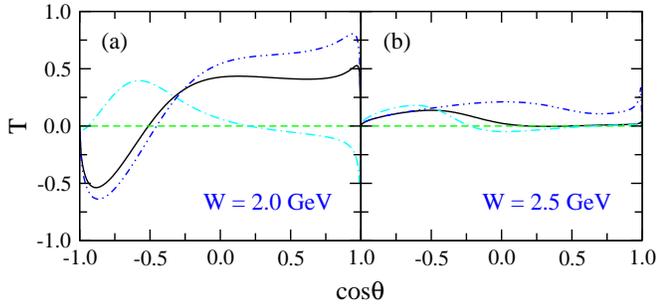}
\caption{(Color online) Same as in Fig.~\ref{fig:beam} for target nucleon asymmetries.}
\label{fig:tagt}
\end{figure}

In Figs.~\ref{fig:beam} and \ref{fig:tagt}, we show the predictions of the photon beam asymmetries ($\Sigma$) and target nucleon asymmetries ($T$) for $\gamma p \to K^+\Lambda(1405)$ at two selected energies in our present model. In these two figures, the black solid lines represent the results from the full calculation; the green dashed, blue dash-double-dotted, and cyan dash-dotted lines represent the results obtained by switching off the individual contributions of the $s$-channel $N(2000){5/2}^+$ exchange, $t$-channel $K$ exchange, and $s$-channel $N$ exchange, respectively, from the full model. 
One sees that for $\Sigma$, all the $N(2000){5/2}^+$, $K$, and $N$ exchanges have significant contributions at $W=2.0$ GeV. At $W=2.5$ GeV, the shape of $\Sigma$ changes dramatically only when the $K$ exchange is switched off. For $T$, most important contributions come from the $N(2000){5/2}^+$ and $N$ exchanges. Note that without the $N(2000){5/2}^+$ exchange, one gets zero target asymmetries, which is not surprising as the amplitudes from the $t$-channel $K$ and $K^\ast$ exchanges, $s$-channel $N$ exchange, $u$-channel $\Sigma$, $\Lambda$, and $\Lambda(1405)$ exchanges, and interaction current are all purely real. It should be mentioned that in effective Lagrangian approaches, usually the widths of $t$-channel and $u$-channel exchanged hadrons are not taken into account in their propagators since the investigated energy region is in the unphysical area of the $t$-channel and $u$-channel reactions. We hope that the spin observables including $\Sigma$ and $T$ can be measured in experiments in the near future, which can help to further constrain the theoretical models and thus lead to a better understanding of the reaction mechanisms and a more reliable extraction of the resonance contents and parameters in this reaction.

\section{Summary and conclusion}  \label{sec:summary}

In the present work, we employ an effective Lagrangian approach at the tree-level approximation to analyze the most recent differential cross-section data from the CLAS Collaboration for the $\gamma p \to K^+\Lambda(1405)$ reaction. To reproduce the data, we consider the $t$-channel $K$ and $K^\ast$ exchanges, $s$-channel $N$ exchange, $u$-channel $\Sigma$, $\Lambda$, and $\Lambda(1405)$ exchanges, and generalized contact current in the background contributions, and take into account the exchanges of a minimum number of nucleon resonances in the $s$ channel in constructing the reaction amplitudes. The full photoproduction amplitudes satisfy the generalized WTI and thus are fully gauge invariant. 

It is found that by introducing in the $s$ channel the exchange of the $N(2000){5/2}^+$ resonance, which is rated as a two-star resonance in the most recent PDG \cite{PDG}, we can achieve a very good description of the available differential cross-section data from the CLAS Collaboration \cite{data2013} for $\gamma p \to K^+\Lambda(1405)$. The predicated total cross sections from the present work are also in qualitative agreement with the corresponding data. It is shown that the cross sections at high energies are dominated by the $t$-channel $K$ exchange. In the near-threshold energy region, significant contributions are seen from the $s$-channel $N$ and $N(2000){5/2}^+$ exchanges, and the $N(2000){5/2}^+$ resonance exchange is responsible for the bump structure exhibited by the total cross-section data around $W\approx 2.1$ GeV. The coupled-channel effects from the intermediate $K^\ast\Sigma$ states to the $\gamma p \to K^+\Lambda(1405)$ reaction will be further investigated in our next-step work.

The predictions of the photon beam asymmetries ($\Sigma$) and target nucleon asymmetries ($T$) from our present theoretical model for the $\gamma p \to K^+\Lambda(1405)$ reaction are presented. These spin observables are expected to be measured in the future experiment, which can be used to further constrain the theoretical models and help to get a more accurate understanding of the reaction mechanism and a more reliable extraction of the resonance contents and parameters in this reaction.

\begin{acknowledgments}
This work is partially supported by the National Natural Science Foundation of China under Grants No.~11475181 and No.~11635009, the Fundamental Research Funds for the Central Universities, and the Key Research Program of Frontier Sciences of the Chinese Academy of Sciences under Grant No.~Y7292610K1.
\end{acknowledgments}

\end{document}